\documentclass[preprint,aip,apl]{revtex4-1}

\usepackage{amsmath}    
\usepackage{graphicx}   
\usepackage{verbatim}   
\usepackage{color}      
\usepackage[hidelinks]{hyperref}   
\newcommand{\Neel}{N\'{e}el}

\begin{document}

\title{Scanning electron microscopy with polarization analysis for multilayered chiral spin textures}
\author{Juriaan Lucassen}
\email{j.lucassen@tue.nl}
\affiliation{Department of Applied Physics, Eindhoven University of Technology, 5600 MB Eindhoven, the Netherlands}
\author{Fabian Kloodt-Twesten}
\affiliation{Institut f\"{u}r Nanostruktur-und Festk\"{o}rperphysik, Universit\"{a}t Hamburg, Jungiusstra{\ss}e 11, 20355 Hamburg, Germany}
\author{Robert Fr\"{o}mter}
\author{Hans Peter Oepen}
\affiliation{Institut f\"{u}r Nanostruktur-und Festk\"{o}rperphysik, Universit\"{a}t Hamburg, Jungiusstra{\ss}e 11, 20355 Hamburg, Germany}
\author{Rembert A. Duine}
\affiliation{Department of Applied Physics, Eindhoven University of Technology, 5600 MB Eindhoven, the Netherlands}
\affiliation{Institute for Theoretical Physics, Utrecht University, Leuvenlaan 4, 3584 CE Utrecht, the Netherlands}
\author{Henk J.M. Swagten}
\author{Bert Koopmans}
\affiliation{Department of Applied Physics, Eindhoven University of Technology, 5600 MB Eindhoven, the Netherlands}
\author{Reinoud Lavrijsen}
\affiliation{Department of Applied Physics, Eindhoven University of Technology, 5600 MB Eindhoven, the Netherlands}

\date{\today}

\begin{abstract}
We show that scanning electron microscopy with polarization analysis (SEMPA) that is sensitive to both in-plane magnetization components can be used to image the out-of-plane magnetized multi-domain state in multilayered chiral spin textures. By depositing a thin layer of Fe on top of the multilayer we image the underlying out-of-plane domain state through the mapping of its stray fields in the Fe. We also demonstrate that SEMPA can be used to image the domain wall chirality in these systems after milling away the capping layer and imaging the topmost magnetic layer directly.
\end{abstract}

\maketitle

Since the observation of room-temperature magnetic skyrmions in thin-film multilayer systems~\cite{Luchaire_skyrmion,Boulle2016,Woo2016}, much progress has been made in understanding the role of the Dzyaloshinksii-Moriya interaction (DMI) in these systems.  However, to further our understanding high resolution imaging techniques are needed that are able to resolve the nanoscale spin texture. Until now a few methods have been used to image the magnetic order in these systems. These are X-ray magnetic circular dichroism photoemission electron microscopy (XMCD-PEEM)~\cite{Boulle2016}, magnetic transmission (soft) X-ray microscopy (MTXM)~\cite{Luchaire_skyrmion,Woo2016}, spin-polarized low-energy electron microscopy (SPLEEM)~\cite{ChenSkyrmion}, magnetic force microscopy (MFM)~\cite{2016arXiv160606034S,2016arXiv160901615B}, Lorentz transmission electron microscopy (LTEM)~\cite{Pollard2017} and imaging with nitrogen vacancy (NV)-centres in diamond~\cite{2016arXiv161100673D}. MFM, and NV-centres, however, provide no direct information on the chirality of the domain walls and skyrmions in out-of-plane (OOP) magnetized systems and LTEM and MTXM require transparent samples\cite{Luchaire_skyrmion,Woo2016,Pollard2017}.

Scanning electron microscopy with polarization analysis (SEMPA~\cite{KOIKE1984,Unguris1986,OEPEN1988}) combines a resolution down to $3$~nm~\cite{Koike} with the capability to map both in-plane (IP) magnetization components or one IP and the OOP component simultaneously~\cite{Allenspach,oepen_review}. It has been demonstrated that SEMPA can be used to image the sense of rotation of domain walls in the epitaxial single layer Pt/Co/vacuum system~\cite{HamburgDW}. SEMPA is also an attractive option for studies of magnetization dynamics with the recent advances in time-resolved SEMPA~\cite{TRsempa}. However, in general, SEMPA experiments face two challenges when trying to analyze multilayer systems. First, the high surface sensitivity (penetration depths less than $1$~nm), which requires a milling step to remove the paramagnetic capping layer before measurement~\cite{UNGURIS2001167,*Oepen2005}. Secondly, today's commercially avaible SEMPA systems are sensitive only to the IP magnetization components, which means that OOP domains can only be observed directly with reduced signal-to-noise ratio by tilting the sample with respect to the spin detector~\cite{Froemter2008}.

In this Letter we describe a method in which an IP SEMPA system is used to image OOP domains in capped systems relevant for skyrmion stabilization. By depositing a thin film of IP Fe on top of the capped OOP multilayer structures, we image the OOP domains because the Fe will be polarized in the direction of the stray fields coming from the system underneath. We show that the amount of evaporated Fe is not critical and that this method can be used to image through both $3$~and~$11$~nm Pt capping layers. We validate this method by comparing it to MFM measurements on the same samples. 

With SEMPA we are able to go one step further; it is also possible to image both the domains and domain walls by mapping both the IP domain wall magnetization direction and OOP domains simultaneously. For the latter experiments the capping layer is removed by ion beam milling, after which a thin layer of Co is deposited to enhance the SEMPA contrast. For the imaging the sample is tilted which gives both IP and OOP contrast. Using this approach, we show that an Ir/Co/Pt multilayer repeat system has clockwise~\Neel~walls at the top surface, which demonstrates that SEMPA can be used to investigate nanoscale multilayered chiral spin textures

The systems which are investigated are Ir/Co/Pt multilayers with a varying number of repeats and thicknesses, as these are the typical material stacks in which skyrmions have been found~\cite{Luchaire_skyrmion,2016arXiv160606034S,2016arXiv160901615B}. We chose thicknesses and repeats for which an as-grown OOP multi-domain state is present to ensure that no field sequences are needed before a SEMPA measurement. The samples are DC magnetron sputtered using Ar at $1\times 10^{-2}$~mbar on a Si substrate with a native oxide in a system with a base pressure of $3\times 10^{-8}$~mbar. The sample compositions are //Ta($4$)/Pt($2$)/X/Pt($2$), with X for the individual samples given by:
\begin{description}
\item[Sample A]{[Pt($1$)/Co($0.9$)/Ir($1$)]$\times 15$}
\item[Sample B]{[Pt($1$)/Co($1$)/Ir($1$)]$\times 15$ (for this sample the Pt capping layer was $10$~nm thick)}
\item[Sample C]{[Pt($1$)/Co($1.2$)/Ir($1$)]$\times 25$}
\item[Sample D]{[Pt($1$)/Co($1.3$)/Pt($1$)]$\times 25$}
\item[Sample E]{[Ir($1$)/Co($1.2$)/Pt($1$)]$\times 25$},
\end{description}
where the thicknesses in parentheses are given in nm. The SEMPA system at the University of Hamburg is described elsewhere~\cite{HighResSemFrom}. Fe and Co overlayers are evaporated directly in the SEMPA chamber using e-beam evaporation. All measurements are performed in the virgin state and at room temperature, except for sample A. It shows no domains in the virgin state and is demagnetized using an oscillating exponentially decaying in-plane field prior to measurement. The magnetization and anisotropy of the samples are determined using a SQUID-VSM at room temperature. MFM measurements are performed under ambient conditions using a NT-MDT Solver P47H with low-moment magnetic tips (NT-MDT FMG01) using a two-pass technique by recording the phase shift~\cite{MFMref}.

\begin{figure}
\centering
\includegraphics{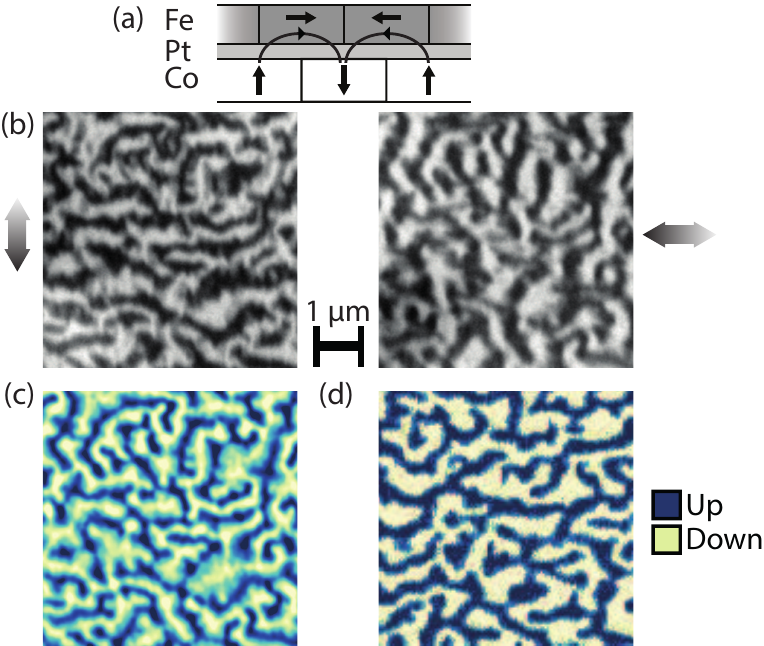}
\caption{\label{fig:figure1}(a) Principle of the IP stray field imaging technique. The stray fields from the OOP Co system align the evaporated IP Fe through the capping layer. (b) SEMPA images of sample B with $3.0$~nm of Fe evaporated on top. The left image shows the (in-plane) up-down magnetization, and the right image shows the simultaneously recorded right-left magnetization, where the arrows denote the relationship between the contrast and the magnetization direction. (c) Spatial divergence of the Fe domain pattern from (b) revealing the underlying OOP Co domains. The divergence was calculated after Gaussian smoothing the SEMPA images. (d) MFM image of the same sample (different area).}
\end{figure}

The stray field imaging technique is based on the principle depicted in FIG.~\ref{fig:figure1}a. By evaporating a thin layer of IP Fe on a capped OOP multilayer stack the IP stray fields emanating from this OOP layer can be imaged by mapping the IP Fe domains with SEMPA. This is different compared to the well-known technique where a layer of exchange-coupled Fe is used to enhance the magnetic contrast~\cite{VanZandtFe} because the dipolar coupling dominates here. As OOP domains in the up-direction act as IP field sources, and down-domains as IP field sinks, the OOP domains can be visualized by taking the spatial divergence of the Fe magnetization. The distinction between source and drain also makes it possible to distinguish between the underlying up and down domains.

To demonstrate this technique, in FIG.~\ref{fig:figure1}b we show vectorial IP SEMPA images of sample B, where $3.0$~nm of Fe has been evaported in situ. The sources and drains are found by calculating the divergence as shown in FIG.~\ref{fig:figure1}c, where the characteristic worm-like domain structure of the underlying Co system becomes visible~\cite{Luchaire_skyrmion,Boulle2016,Woo2016,2016arXiv160606034S,2016arXiv160901615B}. This verifies the principle described in FIG.~\ref{fig:figure1}a and demonstrates that we are able to use Fe decoration to image OOP domains with an IP SEMPA system through a Pt capping layer.

\begin{figure}
\centering
\includegraphics{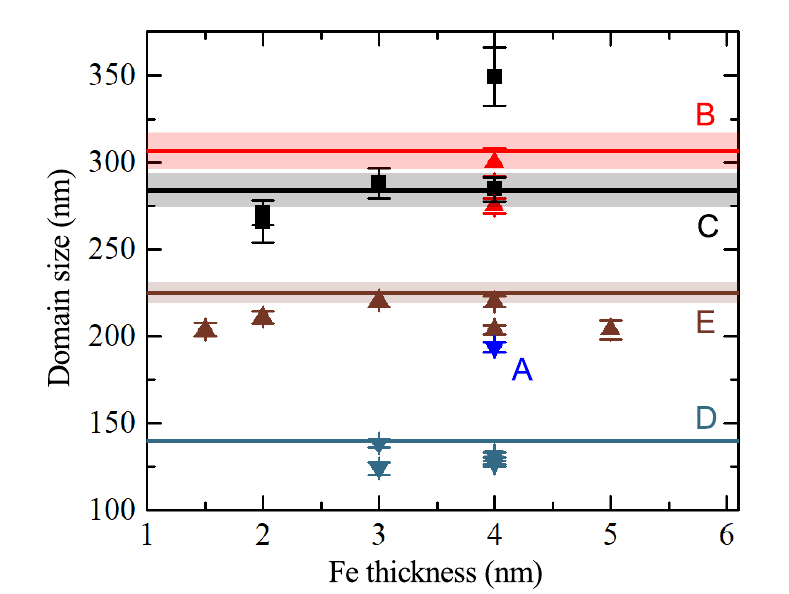}
\caption{\label{fig:comparison} Domain sizes obtained from SEMPA measurements (points) for different evaporated Fe thicknesses. The lines are the domain sizes obtained from MFM measurements, where no Fe was evaporated, and the shaded area indicates the uncertainty of the MFM measurements. The uncertainties given here are the fit uncertainties from the analysis described in the supplementary, but the spread in points at the same Fe thickness suggests the actual uncertainties are larger ($\sim 15$~\% of the domain size).  The labels indicate the corresponding sample. No MFM measurements were performed for sample A.}
\end{figure}

Although it seems highly unlikely that the magnetic domain structure of the Fe is not related to the underlying Co, we further substantiate our claim by comparing SEMPA with MFM imaging. A qualitative comparison is found in FIG.~\ref{fig:figure1}d, where we show a MFM image of the same sample as FIG.~\ref{fig:figure1}c. From this it is clear that the domain structure and size are approximately the same. A more quantitative analysis is given in~FIG.~\ref{fig:comparison} where domain sizes from both SEMPA and MFM measurements are directly compared for all samples investigated. The domain sizes and uncertainties were determined from a quadratic fit to an angular averaged 2D Fourier transform of images such as those depicted in~FIG.~\ref{fig:figure1}c and d (see supplementary material). From several measurements of the domain size on different areas of the same sample, we still find significant variations of the domain size ($\sim 15$\% based upon the $3$~nm Fe data of sample D, and the $4$~nm Fe data of of sample C, D, and E). This indicates that the uncertainty in the analysis is larger than the fit uncertainty, and we attribute this to a large spread in domain sizes that is not properly sampled for the small scan sizes ($25$-$100$~$\mathrm{\mu}$m$^2$) taken.

Based on this analysis we draw two main conclusions. Concerning the SEMPA data alone, we find that there is no discernible change of domain size as the Fe thickness increases. Hence, we may conclude that the evaporated Fe does not impact the magnetic system underneath for the range of thicknesses studied. To further illustrate the fact that the Fe does not impact the system underneath, we find that we can perfectly overlay two images of the exact same area with a $1$~nm difference of evaporated Fe between the two (see supplementary material). Finally, comparing the obtained domain sizes from both SEMPA and MFM we find that they are approximately equal. Although, on average, the MFM domain sizes are a bit larger than the SEMPA domain sizes, both methods agree with each other within the $\sim 15$~\% uncertainty interval.

To image both the domains and domain walls we switch to a different technique. At first, the Pt capping layer is removed using a neutralized Ar ion beam at an acceleration voltage of $150$~eV. The milling is stopped when faint magnetic contrast is obtained. Because we find very little contrast when doing this (quite possibly due to intermixing during growth and/or milling), we also evaporate a small dusting layer of Co that is exchange coupled to the multilayer stack underneath to increase the magnetic contrast in SEMPA. We then tilt the sample with respect to the spin detector, such that the OOP domains appear in the IP magnetization images~\cite{Froemter2008}.

\begin{figure}
\centering
\includegraphics{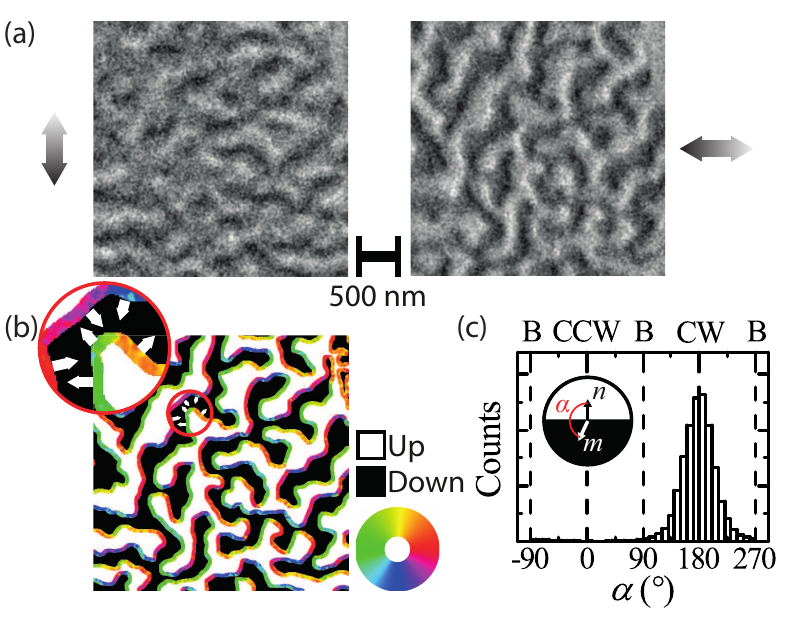}
\caption{\label{fig:figure3}(a) SEMPA images of the [Ir($1$)/Co($1.2$)/Pt($1$)]$\times 25$ (sample E) system with a dusting layer of Co on top. Left image shows the up-down magnetization, and the right image shows the right-left magnetization, where the arrows denote the relationship between the contrast and the magnetization direction. Due to the tilt angle, the bright domains indicate an OOP up domain. (b) Composite image of the results shown in (a). The magnetic domain walls are superimposed on the OOP domains (black/white). The color-wheel indicates the direction of the magnetization in the walls, and the inset is zoomed-in part of the image, where we also denote the magnetization direction by arrows. (c) Histogram of the angle $\alpha$ between the domain wall normal $n$ and the magnetization direction $m$ in the wall for all the pixels in the domain walls shown in (b). The dotted lines indicate the type of wall that corresponds to that $\alpha$, where B indicates a Bloch wall, CW a clockwise~\Neel~wall and CCW a counter-clockwise~\Neel~wall. The inset gives the definitions of $n$, $m$, and $\alpha$.}
\end{figure}
SEMPA images obtained with this method on sample E are shown in FIG.~\ref{fig:figure3}a. The tilt angle during the measurement was $9^{\circ}$ with respect to the spin detector and we observe domain contrast in both IP magnetization images due to this sample tilt~\footnote{The sample is actually tilted in the IP up-down asymmetry axis, but due to a slight misalignment of the spin spin detectors (11$^{\circ}$) and imperfect sample mounting conditions, we actually observe the dominant OOP contrast in the left-right spin direction}. In combination with this OOP domain contrast, we also expect to see IP magnetization contrast due to the domain walls. This is indeed what is observed in the right-left asymmetry image with a darker lining on the left side of the light domains, and brighter lining on the right side of the light domains. These are the magnetization components belonging to the domain walls. Also note that these linings are not present at the top and bottom of the domains, which is a first indication of~\Neel~walls discussed in more detail in the next paragraph. 

In FIG.~\ref{fig:figure3}b both the domain contrast and the magnetization direction in the walls are combined. Here, black and white represent the OOP domain magnetization while the IP magnetization components of the domain walls are shown in color according to the color wheel. The magnetization in the walls is oriented parallel to the domain wall normal (most clearly visible in the inset) and alternates in direction between each successive domain wall which means that we have clockwise (CW) \Neel~walls~\cite{Chen2013,PhysRevB.78.140403}. A more quantitative analysis confirms this and is depicted in FIG.~\ref{fig:figure3}c. Here we plot the histogram of the angle between the domain wall normal and the magnetization direction in the wall for all the measurement pixels in the wall indicated by the colored ribbons (based on the analysis in~Ref.~\onlinecite{Chen2013}). We observe that it is indeed centered around $180^{\circ}$, which implies we have CW \Neel~walls.

Under the assumption that these CW~\Neel~walls are stabilized by the interfacial DM interaction we obtain the sign of $D$ as well as a minimum value for $D$. $D$ is negative because we have CW~\Neel~walls~\cite{PhysRevB.78.140403,0295-5075-100-5-57002}. Using the effective medium approach described in~Ref.~\onlinecite{PhysRevB.95.174423} we calculate the threshold $|D|$ for the formation of complete \Neel~walls. Taking $A=1.6\times 10^{-11}$~$\mathrm{J~m}^{-1}$~\cite{PhysRevLett.99.217208,*Eyrich2014} and $K_{\mathrm{eff}}=0.36$~MJ~m$^{-3}$ as well as $M_{\mathrm{S}}=0.87$~MA~m$^{-1}$ obtained from SQUID-VSM measurements we find $|D|>0.84$~mJ~m$^{-2}$. The sign of $D$ matches theoretical predictions~\cite{PhysRevLett.115.267210,*PhysRevLett.118.219901} for Ir/Co/Pt stacks and corresponds to the sign in inverse Pt/Co/Ir stacks~\cite{doi:10.1021/acs.nanolett.6b01593}, for which $D>0$. The lower boundary for the size also matches literature values, where they find~$|D|=1.7$~mJ~m$^{-2}$ for Pt/Co/Ir~\cite{doi:10.1021/acs.nanolett.6b01593} and~$|D|\sim 0.9$~mJ~m$^{-2}$ for Ir/Co/Pt~\cite{Luchaire_skyrmion,2016arXiv160606034S,2016arXiv160901615B}, where the values have been rescaled such that they match our Co thickness. It should also be possible to extract the actual strength of $D$ by looking at the domain sizes~\cite{Boulle2016,Luchaire_skyrmion,Woo2016}.  However, as detailed in the supplementary material, we encountered several problems when trying to apply this commonly used method to our results.

Although we assumed the domain chirality is fixed by the DMI, we want to mention an effect that is also able to stabilize CW~\Neel~walls at the top interface which is expected to have a significant contribution to the wall structures observed in this paper. For thick OOP layers without DMI, with thicknesses larger than the horizontal Bloch line width $\sqrt{2A/\pi M_{\mathrm{S}}^2}\sim 5.8$~nm (where $M_{\mathrm{S}}=0.87$~MA~m$^{-1}$ the saturation magnetization determined from SQUID-VSM and $A=1.6\times 10^{-11}$~$\mathrm{J~m}^{-1}$~\cite{PhysRevLett.99.217208,*Eyrich2014}) dipolar interactions become important, such that a horizontal bloch line with~\Neel~caps will be formed instead of pure Bloch or \Neel~walls~\cite{HubertDomains,malozemoff1979magnetic,doi:10.1063/1.354964,doi:10.1063/1.3626747}. These walls, driven by flux closure, have a hybrid structure, with CW~\Neel~like walls at the top interface and CCW~\Neel~like walls at the bottom interface, with a Bloch wall in the middle. Based on the analysis from Ref.~\onlinecite{doi:10.1063/1.321822}, we expect hybrid domain walls that lay in between a Bloch and a CW \Neel~wall at the top interface, driven purely by dipolar interactions. This means that dipolar interactions can in part explain the CW chirality of the walls observed here. We stress that the preceding analysis ignores the multilayer structure with the non-magnetic spacers that will reduce the effective exchange interaction~\cite{KAMBERSKY1996301}, reducing the Bloch line width and making this effect even more pronounced. Because this flux closure will affect both the domain wall energy as well as the chirality, it is vital that we understand the role dipolar interactions play in these multilayer systems. 

Lastly, we want to comment on some of the relevant details of the techniques described here, starting with the stray field imaging. In addition to the lack of dependence of the imaged domains on the Fe thickness, we also find that the thickness of the non-magnetic capping layer is not critical. By depositing a thin layer of Fe, we could image through an $11$~nm Pt capping layer (sample B) as well as through several $3$~nm capping layers (samples A, C-E). The theoretical resolution and applicability of this technique depends on several factors. First, the stray fields of the Fe need to be small enough such that the multilayer system remains unaffected. Secondly, the stray fields from the Co need to be large enough to overcome any anisotropy and exchange interaction in the Fe that hinders alignment along the stray fields. In this limit, the resolution of this technique is determined by the domain wall width in the Fe, as this is the ultimate length scale on which the magnetization in the Fe can reverse its direction. Assuming head-to-head transverse walls we find a resolution of $\sim 25$~nm \cite{Nakatani2005750}.

We would also like to point out that this technique is not only applicable to SEMPA, but can likewise be beneficial to other surface sensitive techniques such as SPLEEM~\cite{Spleem} and XMCD-PEEM~\cite{XMCDSPLEEM} if one wants to image OOP domains in capped systems. It is especially attractive for multilayer systems because there is enough magnetic volume such that it is extremely unlikely that a thin layer of Fe will influence the system underneath via stray fields. This makes the technique an extremely valuable addition to the tool-set of imaging magnetic domains (and, potentially, skyrmions) in multilayer systems. For example, we envision the application of this technique to time-resolved SEMPA investigations of skyrmion dynamics. However, note that such an IP capping layer has led to more complex IP domain structures for isolated bubbles~\cite{malozemoff1979magnetic}.

The second technique, where we image the magnetic domain walls directly, is more elaborate. To get the correct domain wall magnetization directions from SEMPA the exposed Co needs to be exchange coupled to the layers underneath. If this is not the case, the chirality of the imaged domain wall will be determined by the DMI of the uncoupled exposed Co layer instead of the DMI of the complete stack. Yet, even though determining the chirality of the underlying stack can be problematic, simple OOP domain imaging using the sample tilt can always be used in multilayer structures due the dipolor and/or exchange coupling between the different magnetic layers. We could also imagine this technique will be very viable for in-situ investigations where the sputtering away of the capping layer is not needed~\cite{HamburgDW}. 

In summary, we have shown that SEMPA is a valuable tool for the imaging of multilayered chiral spin textures. We demonstrate a stray field imaging technique using Fe decoration with which an IP SEMPA is used to image OOP domains. Building on that, we also showed that SEMPA is able to image the domain walls by milling away the capping layer and tilting the sample. This opens up a pathway to fundamental investigations of the domain (wall) structure in chiral spin textures using SEMPA, as well as the option of time-resolved SEMPA in skyrmionic systems~\cite{TRsempa}.

See supplementary materials for (1) a description of the procedure used to extract the domain size; (2) a comparison between two images taken on the same spot with different Fe thicknesses, and (3) a summary of the SQUID-VSM data and domain wall energy calculations.

The authors acknowledge N. S. Kiselev for pointing out that in thick OOP films horizontal Bloch lines are formed. This work is part of the research programme of the Foundation for Fundamental Research on Matter (FOM), which is part of the Netherlands Organisation for Scientific Research (NWO). We gratefully acknowledge funding from Deutsche Forschungsgemeinschaft via Sonderforschungsbereich $668$.
\bibliography{references}
\end{document}